\documentclass[sigconf]{acmart} 

\AtBeginDocument{%
  }

\setcopyright{acmcopyright}
\copyrightyear{2018}
\acmYear{2018}
\acmDOI{XXXXXXX.XXXXXXX}

\acmConference[ICSE 2024]{46th International Conference on Software Engineering}{April 2024}{Lisbon, Portugal}

\usepackage{amsmath}
\usepackage{graphicx}
\usepackage{booktabs}
\usepackage{latexsym} 
\usepackage{array}    
\usepackage{multirow}
\usepackage{subcaption}
\usepackage{algorithm}
\usepackage[noend]{algpseudocode}
\usepackage{threeparttable}
\usepackage{paralist}   
\usepackage{xspace}
\usepackage{color}
\usepackage{adjustbox}
\usepackage{balance}
\usepackage{wasysym}
\usepackage{rotating}   
\usepackage{listings}
\usepackage{tabu}
\usepackage{pifont}
\usepackage{framed}
\usepackage{url}            
\usepackage{hyperref}
\usepackage{makecell}

\usepackage{xpatch}

\usepackage[many]{tcolorbox}

\newif\ifACM
\ACMtrue  

\ifACM
\fi

\ifACM
\newcommand{\myfig}{Figure\xspace}
\else
\newcommand{\myfig}{Fig.\xspace}
\fi

\ifACM
\newcommand{\mysec}{\S}
\else
\newcommand{\mysec}{Sec.\xspace}
\fi

\newcommand{\yq}[1]{{\footnotesize{\textcolor{green}{[YQ: {#1}]}}}\xspace}
\newcommand{\dao}[1]{{\footnotesize{\textcolor{green}{[Dao: {#1}]}}}\xspace}
\newcommand{\review}[1]{{\footnotesize{\textcolor{blue}{[Review: {#1}]}}}\xspace}
\definecolor{cadmiumgreen}{rgb}{0.0, 0.42, 0.24}
\newcommand{\solvedreview}[1]{{\footnotesize{\textcolor{cadmiumgreen}{[\checkmark Review: {#1}]}}}\xspace}

 \renewcommand{\yq}[1]{}
 \renewcommand{\dao}[1]{}
 \renewcommand{\review}[1]{}
 \renewcommand{\solvedreview}[1]{}

\newcommand{\fixme}[1]{{\color{red}{#1}}}
\renewcommand{\fixme}[1]{#1}

\newcommand{\name}{GPTScan\xspace} 
\newcommand{\Cfour}{Code4rena\xspace}
\newcommand{\Eth}{Ethereum\xspace}
\newcommand{\OpenZep}{OpenZeppelin\xspace}

\usepackage[skip=2pt]{caption}
\newcommand{\distance}{3pt}
\DeclareRobustCommand{\answerbox}[2][gray!20]{%
\begin{tcolorbox}[
        breakable,
        left=-2pt,
        right=-2pt,
        top=-4pt,
        bottom=-4pt,
        colback=#1,
        colframe=#1,
        width=\linewidth, 
        enlarge left by=0mm,
        boxsep=5pt,
        arc=0pt,outer arc=0pt,
        ]
        #2
\end{tcolorbox}
}

\usepackage{listings, xcolor}

\definecolor{verylightgray}{rgb}{.97,.97,.97}
\definecolor{codegreen}{rgb}{0,0.55,0}

\lstdefinelanguage{Solidity}{
	keywords=[1]{anonymous, assembly, assert, balance, break, call, callcode, case, catch, class, constant, continue, constructor, contract, debugger, default, delegatecall, delete, do, else, emit, event, experimental, export, external, false, finally, for, function, gas, if, implements, import, in, indexed, instanceof, interface, internal, is, length, library, log0, log1, log2, log3, log4, memory, modifier, new, payable, pragma, private, protected, public, pure, push, require, return, returns, revert, selfdestruct, send, solidity, storage, struct, suicide, super, switch, then, this, throw, transfer, true, try, typeof, using, value, view, while, with, addmod, ecrecover, keccak256, mulmod, ripemd160, sha256, sha3}, 
	keywordstyle=[1]\color{blue}\bfseries,
	keywords=[2]{address, bool, byte, bytes, bytes1, bytes2, bytes3, bytes4, bytes5, bytes6, bytes7, bytes8, bytes9, bytes10, bytes11, bytes12, bytes13, bytes14, bytes15, bytes16, bytes17, bytes18, bytes19, bytes20, bytes21, bytes22, bytes23, bytes24, bytes25, bytes26, bytes27, bytes28, bytes29, bytes30, bytes31, bytes32, enum, int, int8, int16, int24, int32, int40, int48, int56, int64, int72, int80, int88, int96, int104, int112, int120, int128, int136, int144, int152, int160, int168, int176, int184, int192, int200, int208, int216, int224, int232, int240, int248, int256, mapping, string, uint, uint8, uint16, uint24, uint32, uint40, uint48, uint56, uint64, uint72, uint80, uint88, uint96, uint104, uint112, uint120, uint128, uint136, uint144, uint152, uint160, uint168, uint176, uint184, uint192, uint200, uint208, uint216, uint224, uint232, uint240, uint248, uint256, var, void, ether, finney, szabo, wei, days, hours, minutes, seconds, weeks, years},	
	keywordstyle=[2]\color{teal}\bfseries,
	keywords=[3]{block, blockhash, coinbase, difficulty, gaslimit, number, timestamp, msg, data, gas, sender, sig, value, now, tx, gasprice, origin},	
	keywordstyle=[3]\color{violet}\bfseries,
	identifierstyle=\color{black},
	sensitive=false,
	comment=[l]{//},
	morecomment=[s]{/*}{*/},
	commentstyle=\color{codegreen}\ttfamily,
	stringstyle=\color{red}\ttfamily,
	morestring=[b]',
	morestring=[b]"
}

\setcopyright{none}
\renewcommand\footnotetextcopyrightpermission[1]{} 

\begin{document}

\title{When GPT Meets Program Analysis: Intelligent and Effective Detection of Smart Contract Logic Vulnerabilities} 
\title{When GPT Meets Program Analysis: Towards Intelligent Detection of Smart Contract Logic Vulnerabilities in \name} 
\title{\name: Detecting Logic Vulnerabilities in Smart Contracts by Combining GPT with Program Analysis} 


\author{Yuqiang Sun}
\affiliation{%
  \institution{Nanyang Technological University}
  \city{Singapore}
  \country{Singapore}}
\email{suny0056@e.ntu.edu.sg}

\author{Daoyuan Wu}
\authornote{Corresponding author.}
\affiliation{%
  \institution{Nanyang Technological University}
  \city{Singapore}
  \country{Singapore}}
\email{daoyuan.wu@ntu.edu.sg}

\author{Yue Xue}
\affiliation{%
 \institution{MetaTrust Labs}
  \city{Singapore}
  \country{Singapore}}
\email{xueyue@metatrust.io}

\author{Han Liu}
\affiliation{%
  \institution{East China Normal University}
  \city{Shanghai}
  \country{China}}
\email{hanliu@stu.ecnu.edu.cn}

\author{Haijun Wang}
\affiliation{%
  \institution{Xi'an Jiaotong University}
  \city{Xi'an}
  \country{China}}
\email{haijunwang@xjtu.edu.cn}

\author{Zhengzi Xu}
\affiliation{%
  \institution{Nanyang Technological University}
  \city{Singapore}
  \country{Singapore}}
\email{zhengzi.xu@ntu.edu.sg}

\author{Xiaofei Xie}
\affiliation{%
  \institution{Singapore Management University}
  \city{Singapore}
  \country{Singapore}}
\email{xfxie@smu.edu.sg}

\author{Yang Liu}
\affiliation{%
  \institution{Nanyang Technological University}
  \city{Singapore}
  \country{Singapore}}
\email{yangliu@ntu.edu.sg}

\renewcommand{\shortauthors}{Sun et al.}

\begin{abstract}

Smart contracts are prone to various vulnerabilities, leading to substantial financial losses over time.
Current analysis tools mainly target vulnerabilities with fixed control- or data-flow patterns, such as re-entrancy and integer overflow. 
However, a recent study on Web3 security bugs revealed that about 80\% of these bugs cannot be audited by existing tools due to the lack of domain-specific property description and checking.
    Given recent advances in Large Language Models (LLMs), it is worth exploring how Generative Pre-training Transformer (GPT) could aid in detecting \textit{logic vulnerabilities}. 

In this paper, we propose \name, the first tool combining GPT with static analysis for smart contract logic vulnerability detection.
Instead of relying solely on GPT to identify vulnerabilities, which can lead to high false positives and is limited by GPT's pre-trained knowledge, we utilize GPT as a versatile code understanding tool.
By breaking down each logic vulnerability type into \textit{scenarios} and \textit{properties}, \name matches candidate vulnerabilities with GPT.
To enhance accuracy,
\name further instructs GPT to intelligently recognize key variables and statements, which are then validated by static confirmation. 
Evaluation on diverse datasets with around 400 contract projects and 3K Solidity files
shows that \name achieves high precision (over 90\%) for token contracts and acceptable precision (57.14\%) for large projects like Web3Bugs.
It effectively detects ground-truth logic vulnerabilities with a recall of over 70\%, including 9 new vulnerabilities missed by human auditors.
\name is fast and cost-effective, taking an average of 14.39 seconds and 0.01 USD to scan per thousand lines of Solidity code.
Moreover, static confirmation helps \name reduce two-thirds of false positives. 

\end{abstract}

\maketitle

\section{Introduction}
\label{sec:intro}

Smart contracts have emerged as the cornerstone of decentralized finance (DeFi), providing a programmable and automated solution for executing financial transactions.
However, the security of these smart contracts has become a major concern due to various security breaches~\cite{Dao_attack_2016, parity_wallet_hack}.
These breaches have led to financial losses amounting to billions of dollars~\cite{zhou_sok_2023}.
This situation is a disaster for DeFi service providers, posing a significant threat to the entire DeFi ecosystem and the safety of users' assets.

Despite the availability of numerous analysis tools~\cite{noauthor_slither_2023, brent2018vandal, brent2020ethainter, Kalra_Goel_Dhawan_Sharma_2018, tsankov2018securify}, they often focus on vulnerabilities with fixed control- or data-flow patterns, such as re-entrancy~\cite{Sereum19, Clairvoyance20}, integer overflow~\cite{tan2022soltype}, and access control vulnerabilities~\cite{liu2022finding, fang_modifier_issta_2023, ghaleb2023achecker}. 
However, a recent study conducted by Zhang et al.~\cite{zhang_demystifying_2023} on Web3 security bugs reveals that around 80\% of these vulnerabilities remain undetected by existing tools.
These undetected vulnerabilities are primarily associated with the business logic of smart contracts.
Traditional static and dynamic analysis schemes, such as Slither~\cite{noauthor_slither_2023}, do not effectively address these vulnerabilities in smart contracts because they do not aim to comprehend the underlying business logic of smart contracts, nor do they model the functionality or consider the roles of various variables or functions. 
In this paper, we explore how recent advances in Large Language Models (LLMs)~\cite{ChatGPT} or Generative Pre-training Transformer (GPT)~\cite{ouyang_training_2022, kojima2022large} could aid in detecting logic vulnerabilities in smart contracts.
A recent technical report~\cite{David_Zhou_Qin_Song_Cavallaro_Gervais_2023} attempted to use GPT by providing it with high-level vulnerability descriptions for project-wide ``Yes-or-No'' inquiries, which is already easier than typical function-level vulnerability detection.
However, this approach suffered from a high false positive rate of around 96\% and required advanced reasoning capabilities from GPT, necessitating the use of GPT-4 instead of GPT-3.5.
Instead, we treat GPT as a generic and powerful code understanding tool and investigate how this capability can be combined with static analysis to create an intelligent detection system for logic vulnerabilities.

To this end, we propose \name, the first tool that combines GPT with static analysis for detecting logic vulnerabilities in smart contracts.
To leverage GPT's code understanding capability, we break down each logic vulnerability type into code-level scenarios and properties.
\textit{Scenarios} describe the code functionality under which a logic vulnerability could occur, while \textit{properties} explain the vulnerable code attributes or operations.
This approach enables \name to directly match candidate vulnerable functions based on code-level semantics.
However, since GPT-based matching is still coarse-grained, \name further instructs GPT to intelligently recognize key variables and statements, which are then validated by dedicated static confirmation modules.
Moreover, a smart contract project can consist of multiple Solidity files, making it infeasible or costly to directly feed all of them to GPT.
To address this issue, \name employs a multi-dimensional filtering process to effectively narrow down the candidate functions for GPT matching.

We implemented \name with the widely used GPT-3.5-turbo model~\cite{openai_temperature}, which is 20 times more cost-effective~\cite{Pricing} than the advanced GPT-4 model.
Moreover, our multi-dimensional filtering allowed \name to utilize the default 4k context token size instead of 16k, resulting in a more economical solution.
The parameters were mainly kept at their default values, except for the \texttt{temperature} parameter, which was adjusted from the default value of 1 to 0 to reduce the impact of GPT's output randomness.
To further enhance the reliability of GPT's answers and minimize the influence of output randomness, we proposed a trick called ``mimic-in-the-background'' prompting, inspired by the success of zero-shot chain-of-thought prompting~\cite{kojima2022large}. 
For the static analysis part, \name relies on ANTLR~\cite{noauthor_antlr_nodate} and crytic-compiler~\cite{Crytic-compile} to support call graph and data dependency analysis.

To comprehensively evaluate \name under different scenarios, we collected three diverse datasets from real-world smart contracts.
Together, these datasets comprise around 400 contract projects, 3K Solidity files, 472K lines of code, and include 62 ground-truth logic vulnerabilities.
The first dataset, named \textit{Top200}, consists of smart contracts with the top 200 market capitalization. This dataset primarily serves to evaluate the false positive rate of \name.
The second dataset, referred to as \textit{Web3Bugs}, was collected from the recent Web3Bugs dataset~\cite{Zhang_2023}.
The third dataset, called \textit{DefiHacks}, is sourced from the well-known DeFi Hacks dataset~\cite{DefiHacks}, which contains vulnerable contracts that have experienced past attack incidents.
\textit{Top200} and \textit{DefiHacks} primarily comprise cryptocurrency token contract projects, whereas \textit{Web3Bugs} consists of large contract projects audited on the \Cfour platform~\cite{Code4rena}, with an average of 36 Solidity files per project.

\name achieves a low false positive rate of 4.39\% when analyzing non-vulnerable top contracts like \textit{Top200}.
It also demonstrates similar performance in analyzing another set of token contracts, \textit{DefiHacks}, with a precision of 90.91\%.
These results indicate that \name is suitable for massive scanning of on-chain contracts.
Moreover, when analyzing large contract projects in \textit{Web3Bugs}, \name still achieves an acceptable precision of 57.14\%.
Furthermore, \name shows its efficacy in detecting ground-truth logic vulnerabilities in the \textit{Web3Bugs} and \textit{DefiHacks} datasets, with a recall of 83.33\% and an F1 score of 67.8\% for \textit{Web3Bugs}, and a recall of 71.43\% and an F1 score of 80\% for \textit{DefiHacks}.
In particular, \name identifies 9 new vulnerabilities that were not present in the audit reports of \Cfour.
This highlights the value of \name as a useful supplement to human auditors.

A further analysis of \name's running logs reveals that \name is fast and cost-effective, taking an average of only 14.39 seconds and 0.01 USD to scan per thousand lines of Solidity code in the tested datasets.
The relatively higher cost (around 0.018 USD) and slower speed (around 20 seconds) observed for \textit{Web3Bugs} and \textit{DefiHacks} can be attributed to the presence of more complex functions that cannot be filtered out by static filtering and scenario matching.
Furthermore, we diagnose that \name's static confirmation reduces 65.84\% of the original false positive cases in the \textit{Web3Bugs} dataset. 
This finding underscores the importance of combining GPT with static analysis to achieve accurate results.

\textbf{Availability.}
\name has been integrated as a part of MetaScan (\url{https://metatrust.io/metascan}), an industry-leading smart contract security scanning platform~\cite{MetaScanVersion1, MetaScanVersion2}.
Moreover, \name's evaluation data is available at \url{https://sites.google.com/view/gptscan} for facilitating easier comparisons in future work.

\textbf{Roadmap.}
The rest of this paper is organized as follows.
In \mysec\ref{sec:backg}, we introduce some background information.
In \mysec\ref{sec:example}, we motivate the need of both GPT and static analysis.
Following that, in \mysec\ref{sec:tool}, we detail the design of \name, while in \mysec\ref{sec:evaluate}, we evaluate its performance.
We then discuss the applicability and current limitations in \mysec\ref{sec:discuss}.
Finally, we summarize related work in \mysec\ref{sec:related} and conclude in \mysec\ref{sec:conclude}.

\section{Background}
\label{sec:backg}

In this section, we introduce some background about smart contract vulnerabilities and GPT's application in vulnerability detection.

\textbf{Smart contract vulnerability types.}
Smart contracts are self-running programs deployed on blockchain, written in a high-level language called Solidity~\cite{Solidity}.
As described by Zhang et al.~\cite{zhang_demystifying_2023}, there are 26 types of vulnerabilities in smart contracts, categorized into 3 groups.
The vulnerabilities in the first group are hard to exploit, doubtful, or not directly related to the functionalities of a given project.
The second group of vulnerabilities involves the use of simple and general oracles, not requiring an in-depth understanding of the code semantics.
Examples include \textit{Re-entrancy} and \textit{Arithmetic Overflow}.
Such vulnerabilities can be detected by data flow tracing (e.g., Slither~\cite{noauthor_slither_2023}), static symbolic execution (e.g., Solidity SMT Checker~\cite{SMTChecker} and Mythril~\cite{Mythril_2023}) and other static analysis tools~\cite{Mossberg_Manzano_Hennenfent_Groce_Greico_Feist_Brunson_Dinaburg_2019, Kalra_Goel_Dhawan_Sharma_2018, brent2020ethainter}.
The third group of vulnerabilities requires high-level semantical oracles for detection and is closely related to the business logic.
Most of these vulnerabilities are not detectable by existing static analysis tools.
This group comprises six main types of vulnerabilities: (S1) price manipulation, (S2) ID-related violations, (S3) erroneous state updates, (S4) atomicity violation, (S5) privilege escalation, and (S6) erroneous accounting.

\textbf{GPT and its application in vulnerability detection.}
Generative Pre-training Transformer (GPT) models, such as GPT-3.5~\cite{ouyang_training_2022}, are large language models (LLMs) trained on vast text corpora, including source code descriptions of different programming languages and vulnerabilities.
With this knowledge, GPT can understand and interpret source code, enabling zero-shot learning~\cite{kojima2022large}, where examples of vulnerabilities are not needed to detect vulnerabilities in source code.
However, GPT still has a long way to go before it can fully replace humans in code auditing~\cite{gpt4_cannot_beat_human}.
David et al.~\cite{David_Zhou_Qin_Song_Cavallaro_Gervais_2023} provided GPT with vulnerability descriptions and used them to detect vulnerabilities in source code.
They fed the entire project into the GPT-4-32k model to detect 38 types of vulnerabilities in smart contracts.
However, the results were unsatisfactory and even worse than a random model in terms of recall.
Due to the limitations of the GPT model on content length (from 4k tokens in GPT-3.5 to 32k tokens in GPT-4), analyzing complete projects or documents using GPT is not viable, making David et al.'s approach unsuitable for large projects.
Moreover, as GPT has limited logical reasoning capabilities, its results may not always be accurate, necessitating verification using other methods to reduce the false positive rate.

\begin{figure}[!t]
\begin{lstlisting}[
        language=Solidity,
        linewidth=.48\textwidth,
        frame=none,
        xleftmargin=.03\textwidth,
        ]
function deposit(uint256 _amount) external returns (uint256) {
    uint256 _pool = balance();
    uint256 _before = token.balanceOf(address(this));
    token.safeTransferFrom(msg.sender, address(this), _amount);
    uint256 _after = token.balanceOf(address(this));
    _amount = _after.sub(_before); // Additional check for deflationary tokens
    uint256 _shares = 0;
    if (totalSupply() == 0) {
        _shares = _amount;
    } else {
        _shares = (_amount.mul(totalSupply())).div(_pool);
    }
    _mint(msg.sender, _shares);
}
\end{lstlisting}
    \caption{The \textit{Risky First Deposit} (line 8-9) vulnerability.}
    \label{code:eg_first_deposit}
\end{figure}

\begin{figure}[!t]
\begin{lstlisting}[
        language=Solidity,
        linewidth=.48\textwidth,
        frame=none,
        xleftmargin=.03\textwidth,
        ]
function transfer(address account, uint256 amount) external override notPaused returns (bool) {
    require(msg.sender != account, Error.SELF_TRANSFER_NOT_ALLOWED);
    require(balances[msg.sender] >= amount, Error.INSUFFICIENT_BALANCE);
    // Initialize the ILiquidityPool pool variable
    pool.handleLpTokenTransfer(msg.sender, account, amount);
    balances[msg.sender] -= amount;
    balances[account] += amount;
    address lpGauge = currentAddresses[_LP_GAUGE];
    if (lpGauge != address(0)) {
        ILpGauge(lpGauge).userCheckpoint(msg.sender);
        ILpGauge(lpGauge).userCheckpoint(account);
    }
    emit Transfer(msg.sender, account, amount);
    return true;
}
\end{lstlisting}
    \caption{The \textit{Wrong Checkpoint Order} (line 6-7 \& line 10-11).}
    \label{code:eg_wrong_order}
\end{figure}

\begin{figure*}[t!]
\begin{adjustbox}{center}
\includegraphics[width=1.0\linewidth]{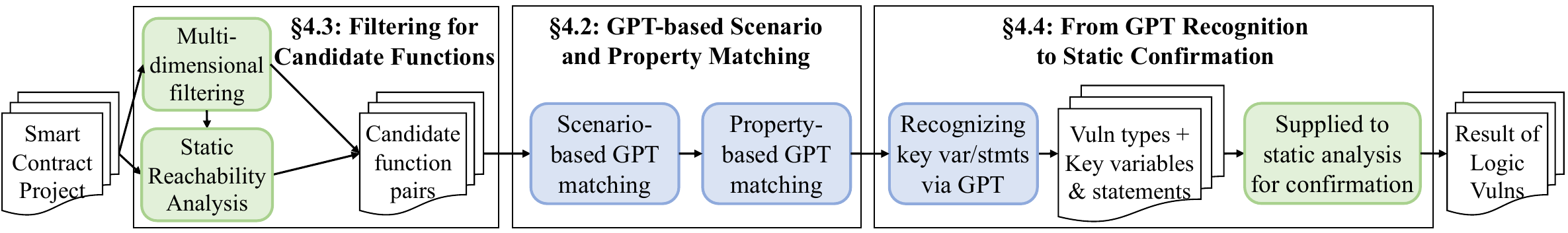}
\end{adjustbox}
\caption{A high-level overview of \name, with blue blocks denoting GPT tasks and green blocks representing static analysis.}
\label{fig:workflow}
\end{figure*}

\section{Motivating Examples}
\label{sec:example}

In this section, we use two real-world smart contract examples to motivate why both GPT and static analysis are needed in the process of detecting logic vulnerabilities.

\textbf{Example 1: Requiring GPT to recognize variables and static analysis to confirm the variable dependency.}
The first example in \myfig~\ref{code:eg_first_deposit} is from the \Cfour~\cite{Code4rena} project \textit{2021-11-yaxis}~\cite{yAxis_2021}.
The vulnerability occurs when the LP (Liquidity Pool~\cite{Lacapra_2023}) token's entire share is minted to the first depositor (line 9) while the current LP token supply is zero (line 8).
Consequently, the first depositor can arbitrarily inflate the price per LP share (e.g., from a small \texttt{\_amount} to an extremely large value; see the detail of an exploit in GitHub issue\cite{yVault_fd_issue}), leading to future token deposits from victim users to be indirectly ``occupied'' by the first depositor.
While static analysis may use hard-coded patterns to detect the \texttt{totalSupply()} logic in line~8,
GPT is necessary to intelligently recognize the variables responsible for holding the deposit amount (\texttt{\_amount}) and the total share of the pool (\texttt{\_shares}) \fixme{to avoid false positives}.
\fixme{Nevertheless}, precisely validating the vulnerable logic from line 8 to 9 falls outside the scope of GPT, making static analysis essential for this task.

\textbf{Example 2: Requiring GPT to recognize statements and static analysis to confirm the statement order.}
The second example in \myfig~\ref{code:eg_wrong_order} is from the Code4Rena project \textit{2022-04-backd}~\cite{Backd_2023}, where the executing order of some statements is incorrect.
The correct order should be to first perform user checkpoints (line 10-11) and then update the balances of the sender and receiver for the transfer (lines 6-7).
Due to this mistake, a user can steal all rewards because the checkpoint is executed after reward transfer~\cite{Backd_2023_issue}.
To detect this vulnerability, GPT is required to understand the semantic of statements and recognize those that perform user checkpoints and those that change user balances.
However, we found that GPT struggles to comprehend the concept of ``before,'' and as a result, relying solely on GPT could report a patched version~\cite{2022-05-backd-StakerVault.sol} of the \texttt{transfer} function as vulnerable.
Static analysis is thus necessary.

\fixme{Based on the above examples, we find that static analysis cannot understand high-level semantic information, and GPT may overlook some low-level information, potentially leading to low recall and high false positives, respectively. Combining these two techniques can complement each other and enhance detection performance.}

\solvedreview{The authors motivate the problem using an argument revealing low recall using existing tools. However, the rest of the paper is focusing on improving precision.}
\dao{We can say more in the response letter. The reviewer does not specifically target at this section; instead, he may challenge our intro. But we assume that existing static analysis tools use domain knowledge to detect those logic vulnerabilities too, then precision problem arises.}

%
%


\section{\name}
\label{sec:tool}

In this section, we present \name's overall design and its three core components from \mysec\ref{sec:overview} to \mysec\ref{sec:confirming}, followed by a summary of some key implementation details in \mysec\ref{sec:implement}.

\subsection{Overview and Challenges}
\label{sec:overview}

\begin{table*}[!t]
    \caption{Breaking down ten common logic vulnerability types into scenarios and properties.}
    \label{tab:rule_sum}
    \scalebox{0.83}{
        \begin{tabular}{|l|l|l|l|}
            \hline
            \textbf{Vulnerability Type}                  & \textbf{Scenario and Property}                                                                                                                                                                                                                                                                                              & \textbf{Filtering Type}    & \textbf{Static Check}     \\ \hline
            \makecell[c]{Approval Not \\Cleared               } & \makecell[l]{\textbf{Scenario: }add or check approval via require/if statements before the token transfer                                 \\ \textbf{Property: }and there is no clear/reset of the approval when the transfer \\finishes its main branch or encounters exceptions                                             } & FNI, FCCE                & VC                        \\ \hline
            \makecell[c]{Risky First \\Deposit             } & \makecell[l]{\textbf{Scenario: }deposit/mint/add the liquidity pool/amount/share                                                          \\ \textbf{Property: }and set the total share to the number of first deposit when \\the supply/liquidity is 0                                                                         } & FCCE                     & DF, VC                    \\ \hline
            \makecell[c]{Price Manipulation \\by AMM          } & \makecell[l]{\textbf{Scenario: }have code statements that get or calculate LP token's value/price                      \\ \textbf{Property: }based on the market reserves/AMMprice/exchangeRate OR the \\custom token balanceOf/totalSupply/amount/liquidity calculation                                   } & FNK, FCCE                  & DF                        \\ \hline
            \makecell[c]{Price Manipulation \\by Buying Tokens} & \makecell[l]{\textbf{Scenario: }buy some tokens                                                                                           \\ \textbf{Property: }using Uniswap/PancakeSwap APIs                                                                                                                                } & FNK, FCE                  & FA                        \\ \hline
            \makecell[c]{Vote Manipulation \\by Flashloan     } & \makecell[l]{\textbf{Scenario: }calculate vote amount/number                                                                              \\ \textbf{Property: }and this vote amount/number is from a vote weight that might \\be manipulated by flashloan                                                                      } & FCCE                     & DF                        \\ \hline
            \makecell[c]{Front Running                        } & \makecell[l]{\textbf{Scenario: }mint or vest or collect token/liquidity/earning and assign them to \\the address recipient or to variable \\ \textbf{Property: }and this operation could be front run to benefit the account/address \\that can be controlled by the parameter and has no sender check in the function code   } & FNK, FPNC, FPT, FCNE, FNM   & FA                        \\ \hline
            \makecell[c]{Wrong Interest \\Rate Order            } & \makecell[l]{\textbf{Scenario: }have inside code statements that update/accrue interest/exchange rate                                     \\ \textbf{Property: }and have inside code statements that calculate/assign/distribute the \\balance/share/stake/fee/loan/reward                                                    } & FCE, CEN                   & OC                        \\ \hline
            \makecell[c]{Wrong \\Checkpoint Order  } & \makecell[l]{\textbf{Scenario: }have inside code statements that invoke user checkpoint                                                   \\ \textbf{Property: }and have inside code statements that calculate/assign/distribute the \\balance/share/stake/fee/loan/reward                                                    } & FCE, CEN                   & OC                        \\ \hline
            \makecell[c]{Slippage                             } & \makecell[l]{\textbf{Scenario: }involve calculating swap/liquidity or adding liquidity, and there is \\asset exchanges or price queries   \\ \textbf{Property: }but this operation could be attacked by Slippage/Sandwich Attack due to no \\slip limit/minimum value check                                                   } & FCCE, FCNCE               & VC                        \\ \hline
            \makecell[c]{Unauthorized \\Transfer              } & \makecell[l]{\textbf{Scenario: }involve transfering token from an address different from message sender                                   \\ \textbf{Property: }and there is no check of allowance/approval from the address owner                                                                                            } & FNK, FCNE, FCE, FCNCE, FPNC   & VC                        \\ \hline
        \end{tabular}
    }
\end{table*}

\myfig~\ref{fig:workflow} illustrates \name's high-level workflow, with blue blocks denoting GPT tasks and green blocks representing static analysis.
Given a smart contract project, which could be a standalone Solidity file or a framework-based contract project containing multiple Solidity files,
\name first performs contract parsing, call graph analysis to determine function reachability, and comprehensive filtering to extract candidate functions and their corresponding context functions.
\name then utilizes GPT to match the candidate functions with pre-abstracted scenarios and properties of relevant vulnerability types.
For the matched functions, \name further recognizes their key variables and statements via GPT, which are subsequently passed to specialized static analysis modules for vulnerability confirmation.

During this three-step process, we need to address the following three challenges:

\noindent \textbf{C1:} A smart contract project may contain tens of Solidity files\footnote{According to our evaluation in \mysec\ref{sec:evaluate}, a \Cfour project has 36 Solidity files on average. In contrast to a recent study~\cite{David_Zhou_Qin_Song_Cavallaro_Gervais_2023}, which claimed to feed entire contracts to the GPT-4 model with 32k tokens, we cannot feed the entire project into the model for analysis.}, making it infeasible or costly to directly feed all of them to GPT. Moreover, the presence of non-vulnerable functions may affect GPT's recognition of vulnerable ones. Therefore, \textit{how to effectively narrow down the candidate functions for GPT matching becomes essential}.

\noindent \textbf{C2:} Existing GPT-based vulnerability detection works~\cite{gpt4_cannot_beat_human, David_Zhou_Qin_Song_Cavallaro_Gervais_2023, Deng_Xia_Peng_Yang_Zhang_2023} typically feed GPT with high-level vulnerability descriptions for vulnerability matching, which either demands advanced reasoning capabilities from GPT or relies on the pre-trained vulnerability knowledge of GPT models. Hence, \textit{can we break down vulnerability types in a manner that allows GPT, as a generic and intelligent code understanding tool, to recognize them directly from code-level semantics?}

\noindent \textbf{C3:} Considering that GPT may produce unreliable answers or fail to recognize differences in similar functions, \textit{further confirming the matched potential vulnerabilities becomes critical}.

Since challenge C1 and C3 are both related to challenge C2, we first present how we tackle C2 in \mysec\ref{sec:matching}, followed by our solutions to C1 and C3 in \mysec\ref{sec:filtering} and \mysec\ref{sec:confirming}, respectively. 

\subsection{GPT-based Scenario and Property Matching}
\label{sec:matching}

Existing GPT-based vulnerability detection works~\cite{gpt4_cannot_beat_human, David_Zhou_Qin_Song_Cavallaro_Gervais_2023, Deng_Xia_Peng_Yang_Zhang_2023} identify vulnerabilities by simply feeding GPT with high-level vulnerability descriptions, such as the one provided for the Front Running vulnerability: ``\textit{An attack where an attacker observes pending transactions and creates a new transaction with a higher gas price, enabling it to be processed before the observed transaction. This is often done to gain an unfair advantage in decentralized exchanges or other time-sensitive operations.}''~\cite{David_Zhou_Qin_Song_Cavallaro_Gervais_2023}.
However, these descriptions are condensed from root causes rather than code properties, making it challenging for GPT to directly interpret code-level semantics.


\textbf{Breaking down vulnerabilities into scenarios and properties.}
\name adopts a different approach by breaking down vulnerability types into code-level scenarios and properties.
Specifically, we use \textit{scenarios} to describe the code functionality under which a logic vulnerability could occur and \textit{properties} to explain the vulnerable code attributes or operations.
Table~\ref{tab:rule_sum} showcases how we break down ten common logic vulnerability types into scenarios and properties.
These vulnerability types were selected from a recent study~\cite{zhang_demystifying_2023} on smart contract vulnerabilities that require high-level semantic oracles~\cite{Zhang_2023}.
The study summarizes six categories of logic vulnerabilities from S1 to S6 (see \mysec\ref{sec:backg}), and we chose ten representative cases from these categories.
For instance, the Approval Not Cleared vulnerability is from S3, which involves missing state update, and the two wrong order vulnerabilities are from S6, relating to incorrect calculating order.
\fixme{Note that in this paper, we manually broke down ten vulnerability types to precisely describe their code-level attributes. To support more logic vulnerability types in future work, we have figured out a GPT-based approach. This approach employs GPT-4 to automatically extract initial scenario and property sentences from past vulnerability reports, validate them using the original vulnerable code, and iteratively regenerate new sentences until a scenario and property sentence pass the original vulnerability validation. However, while the generation of scenario and property sentences can be automated, the prompt used for GPT recognition, which we will explain in \mysec\ref{sec:confirming}, must be manually crafted for different types of vulnerabilities.}

\solvedreview{Who break down the vulnerabilities into scenario and properties? The authors must include the mechanism and validation of the same.}

\solvedreview{There is no insight about how a new vulnerability can be handled, i.e., given the specification of a new vulnerability, how can it be split into Scenario + Property? How to validate if a given split into Scenario+Property is correct? Is it always possible to perform a split? The prompts provided to GPT to extract variables of interest (like the one in Figure 5) also seem to be specialised to a domain.}

\begin{figure}
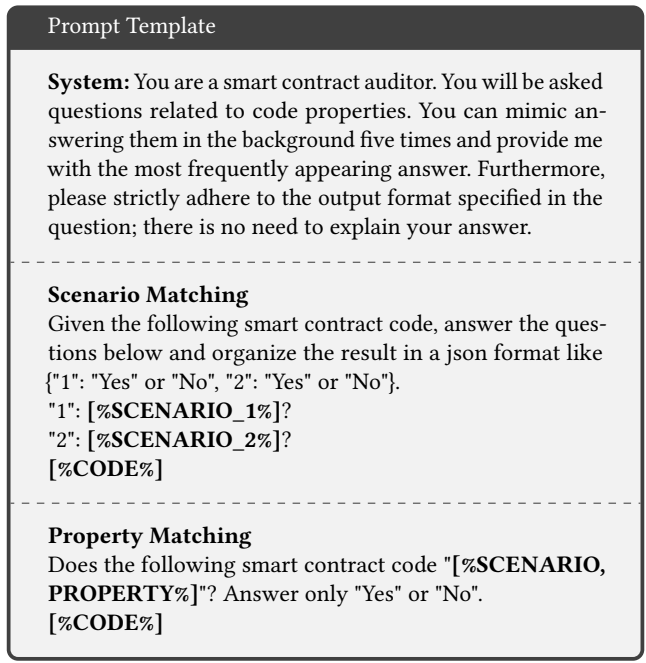

    \begin{tcolorbox}[title=Prompt Template]
        \textbf{System: }You are a smart contract auditor. You will be asked questions related to code properties. You can mimic answering them in the background five times and provide me with the most frequently appearing answer. Furthermore, please strictly adhere to the output format specified in the question; there is no need to explain your answer.
        \tcbline
        \textbf{Scenario Matching}\\
        Given the following smart contract code, answer the questions below and organize the result in a json format like \{"1": "Yes" or "No", "2": "Yes" or "No"\}. \\
        "1": \textbf{[\%SCENARIO\_1\%]}? 

        "2": \textbf{[\%SCENARIO\_2\%]}?\\
        \textbf{[\%CODE\%]}
        
        \tcbline
        \textbf{Property Matching}\\
        Does the following smart contract code "\textbf{[\%SCENARIO, PROPERTY\%]}"? Answer only "Yes" or "No".\\
        \textbf{[\%CODE\%]}
    \end{tcolorbox}
    \caption{Prompt for scenario and property matching.}
    \label{fig:scenarioprompt}
\end{figure}

Each scenario and property can be divided into two parts.
The first part includes a description of the function's functionality, which helps \name perform an initial screening of candidate functions to reduce unnecessary subsequent scanning.
Using \textit{Front Running} as an example, functions affected by this vulnerability type must involve actions like minting, vesting, or transferring tokens of other users.
The approval for such actions is granted in a previous transaction, allowing attackers to front-run the function and gain an unfair advantage.
The second part includes a description of the function's behavior, which is related to the root cause of the vulnerabilities, such as the lack of security checks and incorrect accounting order.
If a function meets the properties of the first part, i.e., scenarios, \name will send the function to GPT again to check if it satisfies both the scenarios and properties. 
If both parts are satisfied, \name considers the function likely to contain a specific type of vulnerability and will confirm it in the later steps.

\textbf{Yes-or-No scenario and property matching.}
With the abstracted scenarios and properties, we utilize them to match candidate functions using GPT.
\myfig~\ref{fig:scenarioprompt} shows the prompt template employed by \name for scenario and property matching, which is designed with three considerations.
Firstly, property matching is performed only for functions that pass our scenario matching.
This separation of scenario and property enables us to query all scenarios in a single prompt, thus saving on GPT costs.
Secondly, during property matching, we double-confirm the scenario with GPT by querying the combination of scenario and property rather than property alone.
Indeed, the scenarios and properties from Table~\ref{tab:rule_sum} are designed to form a complete sentence.
Thirdly, considering that GPT models sometimes provide ambiguous answers or hard-to-parse text, scenario and property matching are designed with yes or no questions only, aiming to minimize the impact of unstructured GPT responses.
Moreover, we instruct GPT to learn the output JSON format for the multiple-choice scenario matching, leveraging GPT's instruction learning capability~\cite{Ouyang_Wu_Jiang_Almeida_Wainwright_Mishkin_Zhang_Agarwal_Slama_Ray}.

\textbf{Minimizing the impact of GPT output randomness.}
Although we use yes-or-no questions to restrict the format of GPT responses, it does not eliminate the inherent randomness of GPT model output.
Consequently, GPT may not provide the same answer for the same question.
To address this, one approach is to set the \texttt{temperature} parameter of GPT models to 0, making the model \fixme{tend to be} deterministic.
\yq{change the cite}
To further enhance the reliability of the answer and minimize the influence of GPT output randomness, we propose a \fixme{trick} called ``mimic-in-the-background'' prompting\fixme{, which} is inspired by \fixme{the successful usage of ``\textit{Let's think step by step.}'' in} the zero-shot chain-of-thought prompting~\cite{kojima2022large} \fixme{-- evaluating such prompting is beyond the scope of this paper.}
As shown in \myfig~\ref{fig:scenarioprompt}, we use a GPT system prompt to instruct the model to mimic answering questions in the background five times and provide the most frequently appearing answer to ensure greater consistency.

\yq{Say that we have weaken the claim.}
\review{The authors proposed a new prompting strategy called "mimic in the background" to enhance the reliability of the answer and lower the randomness. Yet, there was no evaluation about the validity of said approach. Also, how can we guarantee that, indeed, the model did reason five times in the background? It would be more convincing if the model was asked to explicitly provide the five answers in the response and manually extract the most frequent answer. Further, calculating descriptive statistics about these answers would have constituted extra proof of reduced randomness.}

\yq{Answer in the response letter.}
\review{The authors mentions that they are making GPT model "deterministic and repetitive" (lines 505-510). Later it is mentioned that the system asks GPT to mimic answering a query five time to improve the reliability of the answer. If the system is deterministic, how does it help to mimic a query five times?}


\subsection{Multi-dimensional Function Filtering}
\label{sec:filtering}



As mentioned in \mysec\ref{sec:overview}, we need to filter the candidate functions before GPT matching.
Here,
we propose a multi-dimensional filtering to systematically select candidate functions for different vulnerability types.
Moreover, we conduct reachability analysis to retain only the functions that could be accessed by potential attackers.

\textbf{Project-wide file filtering.}
Our multi-dimensional filtering begins with project-wide file filtering, which involves excluding non-Solidity files e..g, those under the ``node\_modules'' directory, test files (e.g., those found in various ``test'' directories), and third-party library files (e.g., those from well-known libraries such as ``openzeppelin'', ``uniswap'', and ``pancakeswap'').
Once these files are filtered out, \name can concentrate on the project's Solidity files themselves.

\textbf{Filtering out \OpenZep functions.}
\OpenZep~\cite{openzeppelin} provides a set of libraries to build secure smart contracts on \Eth, widely used in the smart contract community.
While we have filtered out \OpenZep contracts imported as libraries, we found that \OpenZep functions are often directly copied into many developers' contract code, making our project-wide file filtering ineffective.
To address this, we first perform an offline analysis of \OpenZep's source code to extract all its API function signatures as a whitelist.
Each function signature in the whitelist includes the access control modifier, the class name (sub-contract name), function name, return value types, and parameter types.
For example, the signature of the \texttt{transfer} function in the \texttt{ERC20} contract is \texttt{public ERC20.transfer(address,uint256)}.
Next, \name generates the signature of all candidate functions in the same format and compares them with the signatures in the whitelist.
Note that the signature of the candidate function is generated with both the class name and the name of the inherited class because developers may implement the inherited class.
By conducting this comparison, \name excludes functions with the same signature as those in the whitelist, which we consider secure in this paper.
\fixme{In the future, we will add clone-based filtering that covers function bodies.}

\solvedreview{A function whose signature match OpenZeppelin's function signature is considered secure and ignored. Why is the body of the function not compared? What if the functionality has been overridden?}

\textbf{Vulnerability-specific function filtering.}
After project-wide file and \OpenZep filtering, \name conducts function-level filtering for different vulnerability types, which constitutes the major part of \name's multi-dimensional filtering.
To accommodate various filtering requirements, we have designed a YAML-based~\cite{yaml} filtering rule specification to support the following filtering rules:




\noindent \textbf{FNK:} The Function Name should contain at least one Keyword.

\noindent \textbf{FCE:} The Function Content should contain at least one Expression.

\noindent \textbf{FCNE:} The Function Content should Not contain any Expression.

\noindent \textbf{FCCE:} The Function Content should contain at least one Combination of given Expressions.

\noindent \textbf{FCNCE:} The Function Content should Not contain any Combination of given Expressions.

\noindent \textbf{FPT:} The Function Parameters should match the given Types.

\noindent \textbf{FPNC:} The Function should be Public, and we will Not analyze it with its Caller.

\noindent \textbf{FNM:} The Function should Not contain Modifiers that with access control (e.g., \texttt{onlyOwner}).

\noindent \textbf{CFN:} The Callers of this Function will Not be analyzed.

These filtering rules encompass the basic function name (FNK), the detailed function content (FCE, FCNE, FCCE, and FCNCE), the function parameters (FPT), and the function's caller relation (FPNC, FNM, CFN). Different vulnerabilities will utilize their specific filtering rules.
\fixme{The selection of filters is mainly based on the domain knowledge of the vulnerability types.}
For example, the Risky First Deposit vulnerability shown in \myfig~\ref{code:eg_first_deposit} uses only the FCCE rule type to select any combination of ``total,'' ``supply,'' and ``liquidity,'' either separately or together, \fixme{to ensure that the deposit is related to the calculation of total supply or liquidity of the token.}
On the other hand, \fixme{\textit{Price Manipulation by AMM} is related to the calculation of token prices. In this rule, we used the FNK rule to select functions related to price calculation, and the FCE rule to select functions that contain the keywords ``price,'' ``value,'' and ``liquidity.''} 

\review{The multidimensional filtering stage does not seem to be validated or well rationalized (through logical arguments or citing previous studies). Could those filters have led to the omission of a number of samples that could have impacted the detection performance of GPTScan?}

\solvedreview{No insight is provided about the choice of the filters (FNK, FCE, ...). Why}

\solvedreview{Explain the current methodology and results better – the choice of filters, benchmarks, and the results.}

\textbf{Reachability analysis.}
After filtering, we perform call graph analysis to determine the reachability of candidate functions.
We utilize ANTLR~\cite{noauthor_antlr_nodate}, a lexer and parser generator, to parse the source code of the smart contract project and generate an abstract syntax tree (AST).
Using the AST, we build a call graph for the entire project.
In Solidity, there are four types of access control annotations: \texttt{public}, \texttt{external}, \texttt{internal} and \texttt{private}.
Functions marked as \texttt{public} and \texttt{external} can be called by anyone, making them directly reachable for potential attackers.
Functions marked as \texttt{internal} and \texttt{private} might be called by other reachable functions, so we analyze their reachability and include them if they are reachable.
Moreover, Solidity allows developers to use custom modifiers to perform permission checks before function calls.
For example, functions annotated with \texttt{onlyOwner} are only allowed to be called by the owner, which we consider as unreachable.
Functions that are deemed unreachable are excluded from the subsequent GPT-based matching in \mysec\ref{sec:matching}.

\subsection{From GPT Recognition to Static Confirmation}
\label{sec:confirming}

Although the candidate functions pass the initial filtering and GPT matching on function properties, GPT does not always pay attention to syntactic details, such as conditional statements, require statements, assert statements, revert statements, etc.
A more fine-grained static analysis is necessary to identify potentially vulnerable functions at this stage.
Static analysis tools typically focus on specific variables or statements, while our current inputs are still functions.
This is where we need the assistance of GPT to extract the variables and statements related to the specific business logic described in the prompt.
With these variables and statements, we can use static analysis to confirm whether the vulnerability exists or not.
An example of the prompt sent to GPT to ask for related variables or expressions for \textit{Risky First Deposit} is shown in \myfig~\ref{fig:find_var_prompt}.

\begin{figure}
    \begin{tcolorbox}[title=An Example Prompt for GPT Recognition]

        \textbf{System: } (same as in Figure 4, omitted here for brevity.)
        \tcbline

        In this function, which variable or function holds the total supply/liquidity AND is used by the conditional branch to determine the supply/liquidity is 0? Please answer in a section starts with "VariableB:".\\
        In this function, which variable or function holds the value of the deposit/mint/add amount? Please answer in a section starts with "VariableC:".\\
        Please answer in the following json format: \{"VariableA":\{"Variable name":"Description"\}, "VariableB":\{"Variable name":"Description"\}, "VariableC":\{"Variable name":"Description"\}\}\\
        \textbf{[\%CODE\%]}
    \end{tcolorbox}
    \caption{A prompt for finding related variables/statements.}
    \label{fig:find_var_prompt}
\end{figure}

For each extracted variable or statement, \name instructs GPT to provide a short description.
This description helps determine whether the given variables are relevant to the problem and helps avoid incorrect answers.
If GPT provides variables or statements that do not exist in the context of the function or if the description is not relevant to the question asked, \name terminates the judgment process and considers that the vulnerability does not exist.
On the other hand, if the provided variables and statements pass validation, \name feeds them into a static analysis tool to confirm the existence of the vulnerability using methods such as static data flow tracing and static symbolic execution.
\fixme{Specifically, we have designed the following four major types of static analysis to validate the common logic vulnerabilities listed in Table~\ref{tab:rule_sum}:}


\noindent \textbf{Static Data Flow Tracing (DF):}
This method traces the data flow of variables in the program, where static analysis determines whether the two variables or expressions provided by GPT have data dependencies.
\fixme{For example, \myfig~\ref{code:eg_first_deposit} shows that data flow analysis is needed to determine whether the share is calculated directly with the deposit amount in the \textit{Risky First Deposit} vulnerability.}

\noindent \textbf{Value Comparison Check (VC):}
This method checks whether two variables or expressions are compared in condition statements, such as \textit{require}, \textit{assert}, and \textit{if}.
It is used to ensure that variables or expressions are properly checked before usage.
\fixme{In \textit{Risky First Deposit}, VC is used to check whether the share is compared with the deposit amount. Likewise, in \textit{Unauthorized Transfer}, VC is used to verify whether the sender has been checked before the transfer.}

\noindent \textbf{Order Check (OC):}
This method checks the execution order of two statements, where static analysis determines the order of two statements provided by GPT. 
\fixme{For example, \myfig~\ref{code:eg_wrong_order} shows that OC is used to verify the execution order of performing a transfer and updating the checkpoint in \textit{Wrong Checkpoint Order}.}

\noindent \textbf{Function Call Argument Check (FA):}
This method checks whether an argument of a function call can be controlled by the user or meets specific requirements.
Specifically, GPT provides a function call and the index of an argument, and static analysis determines whether the argument can be controlled by the user or meets the requirements described in the rules.
\fixme{In \textit{Price Manipulation by Buying Tokens}, the function calls need to be checked with FA, as some sensitive variables may be used as parameters and cause price manipulation.} 

\solvedreview{Four type of static analysis have been mentioned, but their usage is not clear. Saying "The analysis ... checks ... does not make it clear how it is used to reduce false positives.}

\subsection{Implementation}
\label{sec:implement}

\name is implemented with 3,640 lines of code (LOC) in Python and 154 LOC in Java/Kotlin.
In this section, we provide a summary of some key implementation details as follows.

\textbf{GPT model and its parameters.}
During the development and testing of \name, we utilized OpenAI's GPT-3.5-turbo model~\cite{openai_temperature}.
Thanks to the multi-dimensional filtering introduced in \mysec\ref{sec:filtering}, \name could use the default 4k context token size instead of 16k, which resulted in a more cost-effective solution.
The parameters were mainly kept at their default values, including \texttt{TopP} set to 1, Frequency Penalty set to 0, and Presence Penalty set to 0.
As discussed in \mysec\ref{sec:matching}, we adjusted the \texttt{temperature} parameter from the default value of 1 to 0 to minimize the impact of GPT output randomness.
During each GPT query, the question is sent with an empty session to ensure that the previous questions and answers do not influence the current question.

\textbf{Static analysis tool support.}
As mentioned in \mysec\ref{sec:filtering}, we utilized ANTLR~\cite{noauthor_antlr_nodate} to parse the Solidity source code and generate an abstract syntax tree (AST).
ANTLR allows for source code analysis without the need for compilation, making it more effective for source code with limited dependencies and build scripts compared to tools relying on compilation, such as Slither~\cite{noauthor_slither_2023}. 
Furthermore, to determine data dependencies between two variables or expressions in \mysec\ref{sec:confirming}, we employed \fixme{a static analysis tool~\cite{MetaScanOpen}} based on the output of crytic-compiler~\cite{Crytic-compile}, a Solidity compiler capable of producing a standard AST for static analysis.
With this approach, we can construct both a control flow graph and a data dependence graph.

\section{Evaluation}
\label{sec:evaluate}

In this section, we conduct experiments to evaluate \name's accuracy, performance, financial overhead, the effectiveness of its static confirmation, and its capability to discover new vulnerabilities.

\textbf{Datasets.}
As shown in Table~\ref{tab:logic_data}, the experiments were conducted on three datasets collected from real-world smart contracts.
These datasets consist of around 400 contract projects, 3K Solidity files, 472K lines of code, and include 62 ground-truth logic vulnerabilities.

The first dataset, called \textit{Top200}, comprises smart contracts with a top 200 market capitalization.
It includes 303 open-source contract projects from six mainstream \Eth-compatible chains~\cite{BlockScope23}.
Since these projects \fixme{are well-audited and widely used}, it is \fixme{reasonable to} assume that they do not contain \fixme{notable} vulnerabilities.
\fixme{This dataset is primarily used to stress-test the false-positive rate of \name in audited contracts.}
The second dataset, called \textit{Web3Bugs,}, was collected from the recent Web3Bugs dataset~\cite{zhang_demystifying_2023, Zhang_2023}, which comprises 100 \Cfour-audited projects.
Among the 100 projects, we included 72 projects that can be directly compiled.
The remaining projects either miss library dependencies or configuration files in their original Web3Bugs repository~\cite{Zhang_2023}.
The third dataset, called \textit{DefiHacks}, come from the well-known DeFi Hacks dataset~\cite{DefiHacks}, which consists of vulnerable token contracts that have incurred past attack incidents.
We included 13 vulnerable projects that certainly cover the vulnerabilities in our ten types. 
\fixme{The ground-truth vulnerabilities in these datasets include those already reported and those newly detected by \name and confirmed by the community.}

\solvedreview{For evaluation, I am unable to understand using a dataset assuming it has 0 vulnerabilities! The assumption statement is : "Since these projects have been deployed on the blockchain for a long time and are very popular, it is assumed that they do not contain vulnerabilities." What is "long time"? What is meant by popular?}

\solvedreview{However, I have concerns about how the ground truth is obtained (or assumed).}

All these projects are compiled with crytic-compiler~\cite{Crytic-compile} using the default configuration.
Note that 17 projects in the \textit{Top200} dataset cannot be compiled with crytic-compiler.
For these projects, \name's static confirmation cannot be applied, and any influenced types of vulnerabilities will be marked as not detected.

\begin{table}[]
    \centering
    \caption{Three diverse datasets for \name's evaluation.}
    \label{tab:logic_data}
    \begin{tabular}{lrrrrr}
    \hline
    Dataset Name & Projects P   & Files F   &  F/P     &    LoC & Vuls \\ \hline 
    Top200       & 303          &   555     &  1.83    & 134,322 &    0 \\        
    Web3Bugs     & 72           & 2,573     &  35.74   & 319,878 &   48 \\        
    DefiHacks    & 13           &    29     &  2.23    & 17,824 &   14 \\ \hline 
    \textbf{Sum} & 388          & 3,157     &  8.14    & 472,024 &   62 \\ \hline
    \end{tabular}
    \vspace{1ex}
\end{table}

\textbf{Research Questions.}
With the datasets above, we aim to answer the following five research questions (RQs): 
\begin{description}
    \item [RQ1:] What is the false positive rate of \name when analyzing a dataset of non-vulnerable top contracts?

    \item [RQ2:] How accurate is \name in analyzing real-word datasets with logic vulnerabilities\fixme{, and how effective is it compared to existing tools}?

    \item [RQ3:] How effective is \name's static confirmation in improving the accuracy of \name?

    \item [RQ4:] What are the running performance and financial costs of \name?

    \item [RQ5:] Can \name discover new vulnerabilities that were previously missed by human auditors?
\end{description}

%
%
%
%

\subsection{RQ1: Measuring False Positives in the Non-vulnerable Top Contracts}
\label{sec:RQ1}

In RQ1, we aim to measure \name's false alarm rate in analyzing non-vulnerable contracts.
This is important because when using \name for massive scanning of on-chain \fixme{token} contracts, we want to minimize the false alarms that require manual checking.

For this purpose, we have collected the \textit{Top200} dataset, which consists of 303 contract projects that are deemed non-vulnerable.
We present \name's analysis result of \textit{Top200} in Table~\ref{tab:logic_accuracy}.
Along with the results of \textit{Web3Bugs} and \textit{DefiHacks}, we calculate the accuracy metrics at the function level for each tested vulnerability type. 
For example, if a project has been tested with five vulnerability types, the sum of all true positives, false positives, true negatives, and false negatives for this project should be 5.
More specifically,





\noindent \textbf{TP} is the number of true positives. One true positive is counted when \name successfully detects a ground-truth vulnerable function for the tested vulnerability type.

\noindent \textbf{TN} is the number of true negatives. One true negative is counted when \name correctly does not report any vulnerable function for the tested vulnerability type.

\noindent \textbf{FP} is the number of false positives. One false positive is counted when \name incorrectly reports one or more vulnerable functions for the tested vulnerability type that has no corresponding ground-truth vulnerabilities in the tested project.

\noindent \textbf{FN} is the number of false negatives. One false negative is counted when \name fails to detect the ground-truth vulnerable function for the tested vulnerability type.

\begin{table}[]
    \centering
    \caption{Overall results of \name's accuracy evaluation.}
    \label{tab:logic_accuracy}
    \begin{tabular}{lrrrrr}
    \hline
    Dataset Name & TP  & TN  & FP  &  FN & Sum \\ \hline
    Top200       &   0 & 283 &  13 &   0 & 296 \\
    Web3Bugs     &  40 & 154 &  30 &   8 & 232 \\
    DefiHacks    &  10 &  19 &   1 &   4 &  34 \\ \hline
    \end{tabular}
\end{table}

Based on the calculation of these metrics, \name reports 13 FPs and 283 TNs for the \textit{Top200} dataset, as shown in Table~\ref{tab:logic_accuracy}.
As a result, the false positive rate of \name in analyzing non-vulnerable top contracts like \textit{Top200} is 4.39\%.
Moreover, we find that \name has a similar precision when analyzing \textit{Top200} and \textit{DefiHacks}, both of which are token contracts with around 2 Solidity files per project (see Table~\ref{tab:logic_data}).
When analyzing large projects like those in \textit{Web3Bugs}, the precision drops from around 90\% (90.91\% for \textit{DefiHacks}) to 60\% (57.14\% for \textit{Web3Bugs}).
The drop in precision is \fixme{likely} because \fixme{the smart contract code} in \textit{Web3Bugs} \fixme{is more diverse, given that \textit{Web3Bugs} contains an average of 36 Solidity files per project} (see Table~\ref{tab:logic_data}).
\fixme{In contrast, smart contracts in \textit{DefiHacks} and \textit{Top200} mainly implement common token functionalities using an average of 2 Solidity files per project, potentially triggering only a limited set of false positives in \name.}
In \mysec\ref{sec:RQ2}, we will further discuss the root causes of \name's false positives.

\solvedreview{"This drop in precision is mainly because each contract project in Web3Bugs has around 36 Solidity files on average (see Table 2), which is 18 times more than in DefiHacks and Top200, increasing the chance of reporting false positives." (lines 798 - 801). I am unable to understand this reasoning. Is there any relationship between the number of lines and false positive? Any studies to support this?}

\solvedreview{Why are we looking at the dataset-specific precision and recall? The first dataset has no TP and FN, implying there is no vulnerability in the ground truth. Looking at them together will reveal that the approach gives us a precision of 53\% (50/(50+44)).}


\answerbox{
\textbf{Answer for RQ1:}
\name achieves a low false positive rate of 4.39\% when analyzing non-vulnerable top contracts like \textit{Top200}.
It also demonstrates similar performance in analyzing 
\textit{DefiHacks}, with a precision of 90.91\%.
These results indicate that \name is suitable for massive scanning of on-chain \fixme{token} contracts.
Moreover, when analyzing large contract projects in \textit{Web3Bugs}, \name still achieves an acceptable precision of 57.14\%.
}


\subsection{RQ2: Efficacy for Detecting Vulnerable Contracts}
\label{sec:RQ2}

In RQ2, we \fixme{assess} the effectiveness of \name in analyzing vulnerable contracts in the \textit{Web3Bugs} and \textit{DefiHacks} datasets\fixme{, and compare its effectiveness with existing tools}.

As shown in Table~\ref{tab:logic_data}, the {Web3Bugs} dataset contains 48 ground-truth logic vulnerabilities, while the \textit{DefiHacks} dataset has 14.
Table~\ref{tab:logic_accuracy} presents the scanning results of these two datasets using \name.
In the case of \textit{Web3Bugs}, \name analyzed a total of 232 vulnerability types across 72 projects, detecting 40 TPs and missing 8 FNs, while incurring 30 FPs.
Consequently, \name achieved a recall of 83.33\% and an F1 score of 67.8\% on this dataset. 
For \textit{DefiHacks}, \name analyzed a total of 34 vulnerability types across 13 projects, detecting 10 TPs and missing 4 FNs, while incurring 1 FP.
On this dataset, \name's recall is 71.43\% and the F1 score is 80\%. 
These results demonstrate that \name effectively detects vulnerable contracts for the covered logic vulnerability types.
Following the initial precision analysis in \mysec\ref{sec:RQ1}, we now analyze the root causes of \name's false negatives and false positives.

In the 12 false negative cases, 4 of them are \textit{Price Manipulation by AMM} and 3 of them are \textit{Risky First Deposit}.
The main reason for these two kinds of false negatives is that \name does not implement an alias analysis in the static check, causing failure during static dataflow tracing.
Additionally, there are 2 cases of \textit{Front Running}, where the scenarios or properties are not accurately matched by GPT.
Furthermore, there are 2 cases of \textit{Slippage} and 1 case of \textit{Unauthorized Transfer}.
Similar to the false positive cases,
The main reason for the false negative \textit{Slippage} cases is the existence of numerous variants of slippage checks, making them challenging to detect using GPT and static analysis.
In the case of \textit{Unauthorized Transfer}, the main reason for this false negative is that GPT failed to distinguish the inconsistency between the comment and code.

\name achieves effective vulnerability detection above at an acceptable false alarm rate.
Among the 44 false positive cases from the three datasets, 15 (34.09\%) were related to \textit{Price Manipulation by AMM}, followed by 11 (25.00\%) cases of \textit{Unauthorized Transfer}.
For these two types, the main reason for the false alarms is that these vulnerabilities require specific triggering conditions involving other related logic, which may not be contained within a single function and its callers or callees.
For example, in \textit{Unauthorized Transfer}, the checks for the allowance/approval from the address owner can occur at various positions in the logic chain and may involve multiple functions.
Similarly, the function that calculates the price with AMM for \textit{Price Manipulation} may not be used by other functions responsible for swapping or buying tokens, leading to the vulnerabilities not being triggered in those circumstances.

Additionally, there were 5 cases of \textit{Risky First Deposit} and 5 cases of \textit{Slippage}.
For \textit{Risky First Deposit}, the false alarms occurred because there were many statements related to checking the supply and setting the share, making it challenging for GPT to understand lengthy code segments accurately.
Regarding \textit{Slippage}, the false alarms were mainly due to two factors.
First, similar to \textit{Unauthorized Transfer}, the check for slippage can happen at any position in the logic chain, and second, slippage checks can take many different forms and variants, making them difficult to detect with GPT and static analysis.
For this vulnerability type, our focus was on achieving a higher recall at the cost of slightly sacrificing precision.
There were also 4 cases of \textit{Wrong Interest Rate Order}, 3 cases of \textit{Approval Not Cleared}, and 1 case of \textit{Wrong Checkpoint Order}.
For \textit{Wrong Interest Rate Order} and \textit{Wrong Checkpoint Order}, these vulnerabilities are intricately related to the business logic of the project itself, making it challenging to reduce false alarms without comprehensive knowledge of the project's design.
As for \textit{Approval Not Cleared}, the false alarms were primarily because the function may not always be used to transfer tokens, causing \name to detect it erroneously.

\fixme{
    \textbf{Comparison with existing tools.}
    While there are many specific static analysis tools (e.g.,~\cite{brent2020ethainter, tsankov2018securify, Mossberg_Manzano_Hennenfent_Groce_Greico_Feist_Brunson_Dinaburg_2019, bose2022sailfish}), they almost do not cover any of the logic vulnerabilities targeted in this paper.
    We thus selected two comprehensive vulnerability detection tools, one open-source tool, Slither~\cite{noauthor_slither_2023}, and MetaScan's online static scanning service~\cite{MetaScanWeb, MetaScanOpen}, referred to as MScan.
    Both tools have over a hundred vulnerability detection rules, but the rules related to \name are \textit{unchecked-transfer}, \textit{arbitrary-send-eth}, and \textit{arbitrary-send-erc20} for Slither (corresponding to \textit{Unauthorized Transfer} in \name), and two \textit{Price Manipulation} vulnerabilities for MScan.

    We ran Slither on all three datasets and found a total of 13,144 warnings.
    Among these, only 101 of \textit{unchecked-transfer}, 23 of \textit{arbitrary-send-eth}, and 22 of \textit{arbitrary-send-erc20} are related to the \textit{Unauthorized Transfer} vulnerability in \name.
    Unfortunately, all of them were false positives after careful manual checking.
    There are mainly two reasons for this. 
    Firstly, Slither does not correlate call chain information. 
    Many false positive cases involve \texttt{internal} or \texttt{private} functions that have already been checked for unauthorized transfer when they are called. 
    In \name, we analyze the current function and its caller together, effectively addressing the issue of missing contextual semantics.
    Secondly, Slither is unable to correctly detect variants of transfer behavior in \textit{Unauthorized Transfer}, such as burning tokens, leading to its inability to detect vulnerabilities in the dataset.
    \name relies on GPT to gain the ability to analyze code semantics, which, when combined with code context and calling relationships, can more accurately address these problems.

    We also ran MScan on the \textit{DefiHacks} dataset, as 12 of the total 14 vulnerabilities in this dataset are related to \textit{Price Manipulation}.
    Among these 12 true \textit{Price Manipulation} vulnerabilities, MScan detected 7, achieving a recall of 58.33\% and a precision of 100\% for \textit{Price Manipulation}.
    However, MScan failed to detect any other type of logic vulnerabilities.
    MScan achieved high precision because it used some attack incidents in the \textit{DefiHacks} dataset to derive hard-coded patterns for \textit{Price Manipulation}, including the matching of specific function and variable names.
    However, in cases where hard-coded patterns are not applicable, MScan cannot generalize to detect variants of \textit{Price Manipulation} vulnerabilities.
}

\solvedreview{The proposed method does not seem to be explicitly validated against previous approaches (either those that are GPT-guided or those solely based on static analysis).}

\solvedreview{The only issue I have with the paper is in its evaluation. The authors themselves cite a lot of existing tools and techniques, including some using GPTs. However, none of the RQs actually compare with any of those. That is really surprising. Comparison with baselines would not only strengthen the paper, but it would also empirically validate the drawbacks of the existing techniques the authors identified earlier in the paper.}

\solvedreview{There is no comparison with other vulnerability detection tools.}

\fixme{
    For GPT-based tools, the only available study at the time of our submission was conducted by David et al.~\cite{David_Zhou_Qin_Song_Cavallaro_Gervais_2023}.
    Unfortunately, they did not release their tool, and there was insufficient information for us to reproduce it.
    Therefore, we rely on the statistics provided in their paper for comparison.
    According to the paper, their pure GPT-based approach achieved a precision of 4.14\%, a recall of 43.84\%, and an F1 score of 7.57\% with the GPT-4-32k model, and a precision of 4.30\%, a recall of 35.62\%, and an F1 score of 7.68\% with the Claude-v1.3-100k model, respectively.
    The false positives are significantly higher than those of \name, mainly because their tool did not validate the GPT output as \name does in \mysec\ref{sec:confirming}, and thus could be more easily affected by GPT's inherent problems like hallucination~\cite{zhang2023sirens}, bias in training data, and ambiguity in questions.
    Indeed, RQ3 in \mysec\ref{sec:RQ3} suggests a similar finding by measuring the GPT-only result in \name (see details in Table~\ref{tab:logic_static}).
}

\solvedreview{Compare with one static analysis based tool (without GPT) and one GPT based tool (without static analysis) each.}

\solvedreview{Explain the result, contrasting why a baseline failed and how GPTScan overcame the shortcomings of the chosen baseline.}

\answerbox{
\textbf{Answer for RQ2:}
\name shows its efficacy in detecting ground-truth logic vulnerabilities in the \textit{Web3Bugs} and \textit{DefiHacks} datasets, with a recall of 83.33\% and an F1 score of 67.8\% for \textit{Web3Bugs}, and a recall of 71.43\% and an F1 score of 80\% for \textit{DefiHacks}\fixme{, better than existing static and GPT-based tools.}

}

\subsection{RQ3: Effectiveness of Static Confirmation}
\label{sec:RQ3}

\begin{table}[t!]
    \caption{Raw functions before and after static confirmation.}
    \label{tab:logic_static}
    \begin{tabular}{lrr}
        \hline
        \textbf{Vulnerability Type} & \textbf{Before} & \textbf{After} \\
        \hline
        Approval Not Cleared & 34 & 12 \\
        Risky First Deposit & 100 & 21 \\
        Price Manipulation by AMM & 187 & 114 \\
        Price Manipulation by Buying Tokens & 8 & 8 \\
        Vote Manipulation by Flashloan & 2 & 0 \\
        Front Running & 6 & 4 \\
        Wrong Interest Rate Order & 150 & 11 \\
        Wrong Checkpoint Order & 49 & 1 \\
        Slippage & 99 & 42 \\
        Unauthorized Transfer & 12 & 8 \\
        \hline
        \textbf{Total} & \textbf{647} & \textbf{221} \\ \hline
    \end{tabular}
\end{table}

In RQ3, we conduct a further analysis of \name's intermediate results on \textit{Web3Bugs} to examine how static confirmation reduces false positives generated by pure GPT-based matching.

Table~\ref{tab:logic_static} shows the raw functions reported by \name before and after static confirmation.
Note that one vulnerability type may have multiple functions (the final result counts either TP or FP once, according to the calculation in \mysec\ref{sec:RQ1}), and these functions are not merged yet (i.e., a function A and the combination of function A and all its callers would be counted multiple times) that will be done in the final result.
Hence, so the number of ``after'' cases shown here is much larger than the final TP+FP in Table~\ref{tab:logic_accuracy}.
From the result, we observe that static confirmation effectively filters out most false positive cases for the vulnerability types: \textit{Wrong Interest Rate Order}, \textit{Wrong Checkpoint Order} and \textit{Risky First Deposit}.
The reason is that the description of scenarios and properties for these three types is coarse-grained, leading to many candidate functions passing the GPT-based matching step.
In static confirmation, \name can further instruct GPT to identify related statements and variables, filtering out those that do not satisfy the vulnerability types.
Overall, after static confirmation, only 221 raw functions remain out of the original 647 functions.
This indicates that static confirmation successfully filters out two-thirds of the false positives.

We further analyze the negative impact of static confirmation.
Among the 426 cases filtered out, only 3 \fixme{ground-truth} cases \fixme{were initially matched by GPT but later excluded by static analysis, resulting in 3 false negatives}.
\fixme{Another false negative} was related to compilation problems.
The remaining four did not pass the GPT-based scenario and property matching step.
This indicates that static confirmation has only a minor impact on the false negatives.

\solvedreview{"[static analysis] having only a minor impact on the false negative cases" (line 947). Please explain. Are you (a) only validating the vulnerabilities reported by GPT or (b) also looking for new vulnerabilities in the static analysis phase? I assume it is (a), but then I am unable to see why it will impact false negatives.}
\answerbox{
\textbf{Answer for RQ3:}
Static confirmation effectively filtered out 65.84\% of the false positive cases in the \textit{Web3Bugs} dataset, while having only a minor impact on the false negative cases.
}

\subsection{RQ4: Performance and Financial Overhead}

\solvedreview{The cost analysis is very trivial. The overall cost could be significantly high (for example, cloud hosting for the web service). I suggest that the authors either carry out a comprehensive cost analysis or change the research question to limit the scope of the cost to the use of OpenAI's API.}

In RQ4, we evaluate the running time and financial costs of \name \fixme{ when using OpenAI's GPT-3.5-turbo API}.
\fixme{We considered only the costs associated with interacting with GPT and conducting static analysis.}
We measured the time and financial cost of \name on all three datasets, and the results are shown in Table~\ref{tab:logic_cost}.
In this experiment, we used tiktoken~\cite{tiktoken_2023}, a tokenization tool published by OpenAI and used for GPT models, to estimate the number of tokens sent and received by \name.
With the number of tokens sent and received, we can estimate the financial cost of \name.
The total number of lines of code is 472K, and it took 6,793.35 seconds and 4.9984 USD to complete the scan.
On average, it takes 14.39 seconds and 0.010589 USD to scan per thousand lines of code.

\begin{table}[t!]
    \caption{Running time and financial costs of \name.}
    \label{tab:logic_cost}
    \begin{tabular}{lrrrrr}
        \hline
        \textbf{Dataset} & \textbf{KL$^{*}$}   & \textbf{T$^{**}$}     & \textbf{C$^{***}$}   & \textbf{T/KL} & \textbf{C/KL}     \\ \hline
        Top200           & 134.32          & 1,437.37          & 0.7507          & 10.70              & 0.005589               \\
        Web3Bugs         & 319.88          & 4,980.57          & 3.9682          & 15.57              & 0.018658               \\
        DefiHacks        & 17.82           &   375.41          & 0.2727          & 21.06              & 0.015303               \\ \hline
        \textbf{Overall} & \textbf{472.02} & \textbf{6,793.35} & \textbf{4.9984} & \textbf{14.39}     & \textbf{0.010589}      \\ \hline
    \end{tabular}
    \leftline{$^{*}$ KL for KLoC; $^{**}$ T for Time; $^{***}$ C for Financial Cost.}
\end{table}

On \textit{Top200}, the scan cost per thousand lines of code is the cheapest, and the scan speed per thousand lines of code is the fastest.
This is because most candidate functions are filtered out in \name's first two steps, without the need for finding related variables and expressions for static check.
On \textit{Web3Bugs} and \textit{DefiHacks}, the scan cost per thousand lines of code is the most expensive and the scan speed per thousand lines of code is the slowest, respectively.
Projects in \textit{Web3Bugs} and \textit{DefiHacks} are more complex than \textit{Top200}, and there are more complex candidate functions to be scanned.
These complex functions could not be filtered by static filtering and scenario matching, which causes more time and financial cost.

\answerbox{
\textbf{Answer for RQ4:}
\name is fast and cost-effective, taking an average of only 14.39 seconds and 0.01 USD to scan per thousand lines of Solidity code in the tested datasets.
The relatively higher cost and slower speed for \textit{Web3Bugs} and \textit{DefiHacks} can be attributed to the presence of more complex functions that cannot be filtered out by static filtering and scenario matching.
}

\subsection{RQ5: Newly Discovered Vulnerabilities}
\label{sec:RQ5}

In RQ5, we perform a thorough analysis of \name's results on the \textit{Web3Bugs} dataset to see if it could identify new vulnerabilities that were previously missed by human auditors.
Interestingly, \name successfully discovered 9 vulnerabilities from 3 different types, which did not appear in the audit reports of \Cfour.
Among these 9 newly discovered vulnerabilities, 5 are \textit{Risky First Deposit}, 3 are \textit{Price Manipulation by AMM}, and 1 is \textit{Front Running}.
In the following paragraphs, we present one example of each type of newly discovered vulnerability for further discussion.

\textbf{Risky First Deposit.}
Among the newly discovered vulnerabilities, 56\% of them are \textit{Risky First Deposit}.
In the example shown in \myfig~\ref{code:fd_89}, on line 11, when the variable \texttt{\_pool} is 0, indicating an empty liquidity pool, the depositor can obtain all the shares from the pool.
The presence of both \texttt{\_totalSupply} and \texttt{\_pool} variables to represent the liquidity amount in the pool may confuse human auditors.
Although lines 5 to 8 properly handle the case when \texttt{\_totalSupply} is 0, this specific condition involving \texttt{\_pool} on line 11 creates a vulnerability that could be missed.

\begin{figure}[t!]
\begin{lstlisting}[
        language=Solidity,
        linewidth=.48\textwidth,
        frame=none,
        xleftmargin=.03\textwidth,
        ]
function deposit(uint _amount) external {
    ...
    uint _pool = balance();
    uint _totalSupply = totalSupply();
    if (_totalSupply == 0 && _pool > 0) { // trading fee accumulated while there were no IF LPs
        vusd.safeTransfer(governance, _pool);
        _pool = 0;
    }
    uint shares = 0;
    if (_pool == 0) {
        shares = _amount;
    } else {
        shares = _amount * _totalSupply / _pool;
    }
    ...
}
\end{lstlisting}
    \caption{\textit{Risky First Deposit} in 2022-02-hubble.}
    \label{code:fd_89}
\end{figure}

\textbf{Price Manipulation by AMM.}
Another 33\% of the newly discovered vulnerabilities are \textit{Price Manipulation by AMM}.
In the example shown in \myfig~\ref{code:flp_28}, the \texttt{pendingRewards} function is used to calculate the rewards that can be claimed by the user.
On line 9, when the pool is not empty, the amount of rewards that can be redeemed by the user is calculated based on the total supply in the pool.
However, the total supply can be controlled by users, allowing them to manipulate the redeemed amount and exploit the contracts.

\begin{figure}[t!]
\begin{lstlisting}[
        language=Solidity,
        linewidth=.48\textwidth,
        frame=none,
        xleftmargin=.03\textwidth,
        ]
function pendingRewards(uint256 _pid, address _user) external view returns (uint256) {
    PoolInfo storage pool = poolInfo[_pid];
    UserInfo storage user = userInfo[_pid][_user];
    uint256 accRewardsPerShare = pool.accRewardsPerShare;
    uint256 lpSupply = pool.lpToken.balanceOf(address(this));
    if (block.number > pool.lastRewardBlock && lpSupply != 0) {
        uint256 multiplier = getMultiplier(pool.lastRewardBlock, block.number);
        uint256 rewardsAccum = multiplier.mul(rewardsPerBlock).mul(pool.allocPoint).div(totalAllocPoint);
        accRewardsPerShare = accRewardsPerShare.add(rewardsAccum.mul(1e12).div(lpSupply));
    }
    return user.amount.mul(accRewardsPerShare).div(1e12).sub(user.rewardDebt);
}
\end{lstlisting}
    \caption{\textit{Price Manipulation by AMM} in 2021-09-sushimiso.}
    \label{code:flp_28}
\end{figure}

\textbf{Front Running.}
There is one case of \textit{Front Running} shown in \myfig~\ref{code:fr_25}, in which the token to be minted should be previously transferred (line 1).
However, anyone can call the \texttt{mint} function to mint tokens that are transferred but not minted, as there is only a check with the cached amount of the contract (line 7), but not the cached amount of a specific user.
This vulnerability allows an attacker to front run the minting process.
When a user has transferred a token but not minted it, the attacker could front run the \texttt{mint} function to mint the token before the legitimate user.

\begin{figure}[t!]
\begin{lstlisting}[
        language=Solidity,
        linewidth=.48\textwidth,
        frame=none,
        xleftmargin=.03\textwidth,
        ]
/// @notice The lp tokens that the user contributes need to have been transferred previously, using a batchable router.
function mint(address to)
    public
    beforeMaturity
    returns (uint256 minted)
{
    uint256 deposit = pool.balanceOf(address(this)) - cached;
    minted = _totalSupply * deposit / cached;
    cached += deposit;
    _mint(to, minted);
}
\end{lstlisting}
    \caption{\textit{Front Running} in 2021-08-yield.}
    \label{code:fr_25}
\end{figure}



\answerbox{
\textbf{Answer for RQ5:}
\name identified 9 new vulnerabilities not present in the audit reports of \Cfour.
This highlights the value of \name as a useful supplement to human auditors.
}



%
%


%

\section{Discussion}
\label{sec:discuss}

In this section, we discuss \fixme{the current limitations in \name and the potential use of employing other GPT models}.


\textbf{Current limitations in design and implementation.}
In \mysec\ref{sec:filtering}, the modifiers filtering part only utilized a whitelist to filter the modifiers with access control.
However, this filtering method can lead to false positives or negatives of vulnerabilities.
To enhance accuracy, a more precise approach is required, which involves retrieving the definition of modifiers and conducting a detailed semantic analysis on them.
For the static analysis part in \mysec\ref{sec:confirming}, a simple method was used to analyze the control flow graph and data dependence graph.
This analysis is not path-sensitive, meaning that some path-related issues, such as the reachability of certain execution paths under specific conditions, might be overlooked.
It could be improved by introducing symbolic execution engines to the static analysis part.

\textbf{The use of other GPT models \fixme{and parameters}.}
As mentioned in \mysec\ref{sec:implement}, \name employs the widely used GPT-3.5-turbo model~\cite{openai_temperature} as its GPT model.
We also conducted a preliminary test using GPT-4, but we did not observe a notable improvement, while the cost increased 20 times. This finding suggests that \name does not necessarily require more powerful GPT models.
\fixme{
    As the temperature parameter is set to zero, the answers of the GPT model tend to be deterministic. 
    A higher temperature might lead to more creative answers, but it could also result in more false positives or false negatives.
    However, reproducing results becomes more challenging with a higher temperature.
}
In the future, we plan to conduct a systematic test of various GPT models within the context of \name, including Google Bard, Claude (when we have API access to them), and the self-trained LLaMA model, \fixme{as well as the influence of different parameters on \name}.
\solvedreview{It is relevant to see the temperature parameter set to zero after a quick analysis. However, the authors may consider adding it to the discussion to reveal the tradeoffs or potential drawbacks.}


\section{Related Work}
\label{sec:related}

In this section, we discuss some related work.
Various research and tools have focused on vulnerability detection in smart contracts.
Traditional static analysis tools, such as Slither~\cite{noauthor_slither_2023}, Vandal~\cite{brent2018vandal}, Ethainter~\cite{brent2020ethainter}, Zues~\cite{Kalra_Goel_Dhawan_Sharma_2018}, and Securify~\cite{tsankov2018securify}, are used to analyze the source code and detect vulnerabilities.
Symbolic execution tools like Manticore~\cite{Mossberg_Manzano_Hennenfent_Groce_Greico_Feist_Brunson_Dinaburg_2019} and Mythril~\cite{Mythril_2023} can perform bound checks and detect vulnerabilities in bytecode and source code.
These analysis tools have been applied to detect vulnerabilities in smart contracts, such as re-entrancy~\cite{Sereum19, Clairvoyance20}, arithmetic overflow~\cite{tan2022soltype}, state inconsistency problems~\cite{bose2022sailfish}, and access control problems~\cite{liu2022finding, fang_modifier_issta_2023, ghaleb2023achecker}.
Dynamic analysis tools, such as fuzz testing~\cite{grieco2020echidna, jiang2018contractfuzzer, wustholz2020harvey, zhang2019mpro}, automatically generate test cases or inputs for smart contracts to find abnormal behaviors during runtime.
Formal verification techniques like Verx~\cite{permenev2020verx} and VeriSmart~\cite{so2020verismart} can be used to check user-provided specifications.
Nevertheless, Zhang et al.~\cite{zhang_demystifying_2023} suggested that more than 80\% of exploitable bugs are machine undetectable.

Before the advent of ChatGPT (GPT-3.5)~\cite{ouyang_training_2022}, most NLP-based vulnerability detection methods~\cite{cheshkov2023evaluation, thapa2022transformer, omar2023detecting, chen2023diversevul, wu2021peculiar} involved feeding code into binary or multi-classification models.
Now, with the development of instructing GPT~\cite{selfinstruct} and other research providing few-shot learning capabilities~\cite{brown2020language}, interactive solutions can be used for tasks like code repair~\cite{xia2023keep, jiang2023impact} and vulnerability detection~\cite{David_Zhou_Qin_Song_Cavallaro_Gervais_2023}. However, according to the research by David et al.~\cite{David_Zhou_Qin_Song_Cavallaro_Gervais_2023}, the GPT-4 model itself cannot accurately detect vulnerabilities.
Chen et al.~\cite{feng2023prompting} fine-tuned the GPT-3 model for improved performance in GUI graphical interface testing tasks and utilized it for automated testing of Android applications. Additionally, PentestGPT~\cite{Deng_2023} and ChatRepair~\cite{xia2023keep} utilized feedback from the execution results to enhance the performance of the GPT model during interactions.

\section{Conclusion}
\label{sec:conclude}

In this paper, we proposed \name, the first tool combining GPT with static analysis for smart contract logic vulnerability detection.
\name utilized GPT to match candidate vulnerable functions based on code-level scenarios and properties, and further instructed GPT to intelligently recognize key variables and statements, which were then validated by static confirmation.
Our evaluation on three diverse datasets with around 400 contract projects and 3K Solidity files showed that \name achieves high precision (over 90\%) for token contracts and acceptable precision (57.14\%) for large projects, as well as a recall of over 70\% for detecting ground-truth logic vulnerabilities.
\name is fast, cost-effective, and capable of discovering new vulnerabilities missed by human auditors.
In future work, we will expand \name's support for more logic vulnerability types.

\section*{Acknowledgements}
We thank Dawei Zhou, Zhe Wang, Guorui Fan, Liwei Tan, Hao Zhang, and other colleagues at MetaTrust Labs for their help with \name, as well as anonymous reviewers for their constructive feedback.
This research/project is supported by the National Research Foundation Singapore and DSO National Laboratories under the AI Singapore Programme (AISG Award No: AISG2-RP-2020-019), the National Research Foundation, Singapore, and the Cyber Security Agency under its National Cybersecurity R\&D Programme (NCRP25-P04-TAICeN). Any opinions, findings and conclusions or recommendations expressed in this material are those of the author(s) and do not reflect the views of National Research Foundation, Singapore and Cyber Security Agency of Singapore.

\bibliographystyle{ACM-Reference-Format}
\bibliography{main}


\end{document}